\documentclass[%
 preprint,
 amsmath,amssymb,
 aps, physrev,
]{revtex4-2}
 
\usepackage{graphicx}
\usepackage{dcolumn}
\usepackage{bm}
\usepackage{subcaption}  %
\usepackage{float}       %
\usepackage{subcaption}  %
\usepackage{float}       %
\usepackage{booktabs}    %
\usepackage{amsmath}
\usepackage{amssymb}
\usepackage{booktabs}    %
\usepackage{xcolor}       %

\begin{document}

\title{\textbf{Energy-Embedded Neural Solvers for One-Dimensional Quantum Systems} }%

\author{Yi-Qiang Wu}
 \affiliation{Faculty of Artificial Intelligence in Education, Central China Normal University,Wuhan, Hubei 430079, China}
\author{Xuan Liu}
\affiliation{College of Science, Wuhan University of Science and Technology, Wuhan, Hubei 430065, China}
\author{Hanlin Li}
\email{lihl@wust.edu.cn}
\address{College of Science, Wuhan University of Science and Technology, Wuhan, Hubei 430065, China}
\author{Fuqiang Wang}
\affiliation{School of Science, Huzhou University, Huzhou, Zhejiang 313000, China}
\author{\thanks{Department of Physics and Astronomy, Purdue University, West Lafayette, Indiana 47907, USA}}
\date{\today}

\begin{abstract}
 Physics-informed neural networks (PINN) have been widely used in computational physics to solve partial differential equations (PDEs). In this study, we propose an energy-embedding-based physics-informed neural network method for solving the one-dimensional time-independent Schr\"{o}dinger equation to obtain ground- and excited-state wave functions, as well as energy eigenvalues by incorporating an embedding layer to generate process-driven data. The method demonstrates high accuracy for several well-known potentials, such as the infinite potential well, harmonic oscillator potential, Woods-Saxon potential, and double-well potential. Further validation shows that the method also performs well in solving the radial Coulomb potential equation, showcasing its adaptability and extensibility. The proposed approach can be extended to solve other partial differential equations beyond the Schr\"{o}dinger equation and holds promise for applications in high-dimensional quantum systems.

\end{abstract}

\maketitle

\section{\label{sec:level1a}Introduction}
The Schr\"{o}dinger equation is fundamental in quantum mechanics, describing quantum systems with wave functions and eigenstate energies, allowing determination of their ground and excited states. This information is essential for understanding the energy-level structures and dynamic behaviors of quantum systems, with extensive applications in quantum chemistry, condensed matter physics, and materials science. For example, it plays a vital role in predicting molecular energy levels, investigating the electronic properties of condensed matter, and developing new materials~\cite{Gonzlez2020QuantumCA}. Despite its fundamental role, the Schr\"{o}dinger equation remains a formidable challenge to solve, especially for the excited states of complex systems. Traditional numerical approaches, such as Density Functional Theory (DFT) and Variational Monte Carlo (VMC), though exceptionally effective for ground-state problems, often fall short when addressing excited-state issues. DFT struggles to describe strongly correlated electronic systems due to the limitations of exchange-correlation functionals~\cite{Higgott2019variationalquantum,PhysRevLett.109.266404,PhysRevLett.114.183002,PhysRevLett.119.263401}. The variational method constructs trial wave functions with unknown parameters to form an energy functional. These parameters are optimized to minimize the functional. However, the method strongly relies on the choice of trial wave functions, limiting its applicability to complex systems.

In recent years, machine learning technologies, exemplified by deep learning, have achieved revolutionary advances in areas such as image recognition and natural language processing~\cite{10.1145/3065386,10.1093/nsr/nwx110,doi:10.1126/science.aab3050,Deep_learning_to_predict_sequence_specificity}. Neural networks, renowned for their robust universal approximation capabilities, efficiently fit complex functions. The universal approximation theorem demonstrates that neural networks, given sufficient hidden layers and neurons, can approximate any continuous function to an arbitrary degree of precision~\cite{G1989Approximation,1989Multilayer}. Neural networks can also model input-output mappings directly, enabling accurate estimations of target functions. As a result, they have become powerful tools for tackling quantum mechanics challenges~\cite{MEHTA20191,10.3389/fphy.2023.1061580,BARRETT2013131,Torlai2018}.

In the realm of neural network research based on the variational principle, the Fermionic Neural Network has emerged as a potent wave function ansatz, demonstrating exceptional prowess in solving the Schr\"{o}dinger equation for multi-electron systems~\cite{PhysRevResearch.2.033429}. Similarly, PauliNet, using the Hartree-Fock solution as a baseline, has shown promise in accurately determining both ground and excited states of multielectron systems~\cite{HAN2019108929,Entwistle2023,liu2023calculateelectronicexcitedstates}. Furthermore, related neural network architectures have also achieved success in tackling the multinucleon Schr\"{o}dinger equation~\cite{PhysRevLett.127.022502}. Although these methods achieve remarkable performance in specific applications, they still face notable limitations. For instance, they rely heavily on the manual selection of trial wave functions, which limits their ability to adapt to different potential energy systems and reduces their accuracy for various excited states. Furthermore, the computational costs associated with the solution of complex systems remain substantial.

To address these challenges, Physics-Informed Neural Networks (PINNs) have emerged as a promising solution~\cite{Yang2018PhysicsInformedGA,Alt2023}. PINNs incorporate physical equations, such as the Schr\"{o}dinger equation, as constraints in the neural network. This ensures that the network learns from the data while adhering to the physical laws. Deep Neural Networks (DNNs) are used to generate initial wave functions, which are then optimized to satisfy partial differential equations(PDEs), initial conditions, and boundary conditions, thereby transforming physical problems~\cite{Alt2023} into optimization problems. This approach has demonstrated outstanding performance not only in solving forward and inverse problems of PDEs, but also in applications involving various types of PDEs, fluid dynamics, and quantum mechanics~\cite{RAISSI2019686,DBLP:journals/corr/abs-1711-10561,DBLP:journals/corr/abs-1711-10566,doi:10.1137/18M1229845,JALILI2024125089,SIRIGNANO20181339,KHARAZMI2021113547,Jagtap2020ExtendedPN,GAO2021110079,9282004,cai2021physicsinformedneuralnetworkspinns,2023-CNPC51}. However, PINNs focus primarily on the loss optimization of result-based data, while lacking the ability to effectively handle process-based data. This limitation makes the training process challenging and hinders the application of PINNs in scenarios that involve complex potentials.

To tackle the previously mentioned problems, this study proposes a PINN solver based on energy embedding. The proposed model employs a generator composed of fully connected layers and residual layers to produce wave functions. Using the universal approximation theorem and residual layers, the model approximates true wave functions with high precision, ensuring accurate predictions for both energy and wave functions. In addition, an embedding layer is incorporated to capture the intrinsic relationship between energy and wave functions, enabling the model to adaptively handle diverse potential systems and excited states. This design substantially enhances the generalization ability of the model while simplifying the training procedure. The architecture is grounded in the fundamental principles of the Schr\"{o}dinger equation and adopts a modular design with low component coupling. This ensures excellent scalability and adaptability, making the approach applicable to other PDEs with similar characteristics.

\section{\label{sec:level1b}Methods}
\begin{figure}[h]
    \centering
    \includegraphics[width=1.0\textwidth, page=1]{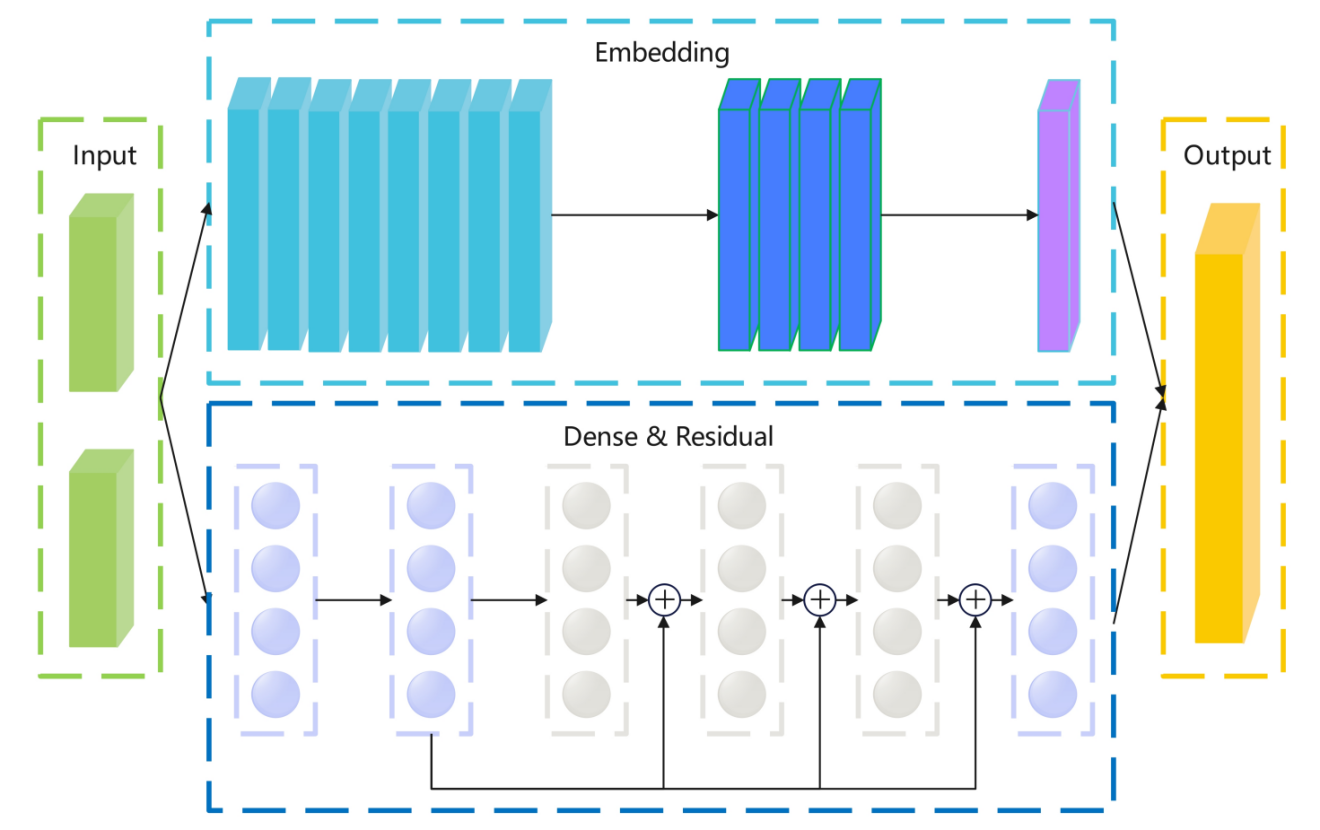}
    \caption{Proposed neural network architecture for solving quantum systems. The embedding layer handles energy representation, while dense and residual layers generate the wavefunctions.}
    \label{fig:DNN}
\end{figure}

In this study, we aim to solve the one-dimensional time-independent Schr\"{o}dinger equation, which can be formulated as:
\begin{equation}
    -\frac{\hbar^2}{2m} \frac{d^2\psi(x)}{dx^2} + V(x)\psi(x) = E\psi(x)
\end{equation}
where $\hbar$ is the reduced Planck constant, $m$ is the particle mass, $V(x)$ represents the potential energy, $\psi(x)$ is the wave function, and $E$ is the corresponding energy eigenvalue. Our goal is to design a unified neural network architecture capable of adapting to different $V(x)$ functions and computing multiple eigenstates and eigenenergies. Figure \ref{fig:DNN} illustrates our neural network architecture.

In the input layer, the coordinate $x$ is discretized into $M$ equally spaced points within the defined domain, with a spacing of $\Delta x$. The discretized coordinates $\mathbf{X}^M=\{x_1,x_2,\dots,x_M\}$ and their corresponding potential values $\mathbf{V}^M=\{V_1,V_2,\dots,V_M\}$ are combined into a $\mathbf{R}^{M \times 2}$ matrix, which serves as the network input. To adapt to different potential systems, we incorporate an embedding layer. This layer calculates embedding vectors for various energy levels under the given potential. Based on the boundary conditions of the physical system, orthogonality constraints, and energy minimization conditions, the embedding layer outputs the embedding vectors $\mathbf{P}^{N \times M}$ corresponding to each energy state. Finally, using the one-hot vector representation of the current eigenstate, the embedding layer produces the specific embedding vector $\mathbf{P}^{M}$. 

We utilize a set of generators to produce wave functions $\psi$. Each generator consists of six fully connected layers and three residual layers. The neural network maps the spatial position $x$ to the corresponding wave function $\psi$ through the respective generator, which applies nonlinear transformations. By leveraging the automatic differentiation capabilities of neural networks, we efficiently compute the first ($f_P$) and second ($f_{PP}$) derivatives of the wave function.

The loss function is designed based on the physical constraints of the Schr\"{o}dinger equation. It is solely dependent on the intrinsic properties of the Schr\"{o}dinger equation, ensuring the scalability of the network. The training process is primarily conducted in an unsupervised manner.

\subsection{\label{sec:level2a}Energy Embedding}
\subsubsection{\textbf{Energy Transformation}}
The input layer represents the discrete coordinates $\mathbf{X}^M$ and potential energy values $\mathbf{V}^M$ as a matrix. For simplicity, we set $\hbar=1$ and $m=1$, and the wave function $\mathbf{P}^M$ satisfies the following Schr\"{o}dinger equation:

\begin{equation}
    - \frac{P_{i+2}+P_{i}-2P_{i+1}}{2(\Delta x)^2} + V_iP_i=eP_i
\end{equation}

Rewriting the above, we obtain:
\begin{equation}
    P_{i+2}=2P_{i+1}+(2(\Delta x)^2(V_i-e)-1)P_i
\end{equation}
We define the vector $\mathbf{B}^{2\times 1} = \{P_2, P_3\}$ and the matrix $\mathbf{H}^{M \times 2}$, such that:

\begin{equation}
    \mathbf{P}^M=\mathbf{H}^{M\times 2} \times \mathbf{B}^{2\times 1}
\end{equation}

Given an energy $e$, let $a_i = 2(\Delta x)^2 (V_i - e) - 1$, then:
\begin{equation}
    H_{i+2}=2H_{i+1}+a_iH_{i}
\end{equation}
Using this recurrence relation, the matrix $\mathbf{H}^{M \times 2}$ can be calculated as:
\begin{equation}
    \mathbf{H}^{M \times 2} = \begin{bmatrix}
    \frac{-2}{a_1} & \frac{1}{a_1} \\
    1 & 0 \\
    0 & 1 \\
    a_2 & 2 \\
    \vdots & \vdots
    \end{bmatrix}
\end{equation}
where $P_M = H_{M,0} P_2 + H_{M,1} P_1$. From the boundary conditions $P_M = 0$, we deduce $P_1=-\frac{H_{M,0}}{H_{M,1}}P_2$. We set $P_2=0.001$. Since the wave function will be normalized in subsequent steps, the specific value of $P_2$ does not affect the final result. This gives us the unnormalized $\mathbf{P}^M$.

To obtain the normalized $\mathbf{P}^M$, we use numerical integration for normalization:

\begin{equation}
    \sum_{i=1}^{M} (P_i)^2 \Delta x \approx 1
\end{equation}

Thus, the normalized wave function is given by:

\begin{equation}
    P^M \leftarrow \frac{P^M}{\sqrt{\sum_{i=1}^{M} (P_i)^2 \Delta x}}
\end{equation}

For energy, we define a transformation function $TransE$, which maps an energy value $e$ to the normalized wave function vector $\mathbf{P}^M$. For a set of energy levels $\mathbf{E}^K = \{e_1, e_2, e_3, \dots, e_K\}$, we have:
\begin{equation}
    TransE(\mathbf{E}^K) = \mathbf{P}^{K \times M}
\end{equation}

\subsubsection{\textbf{Embedding transformation matrix}}
In the embedding process, each eigenstate is represented as a one-hot vector $\mathbf{S}^N$, where $N$ is the total number of eigenstates. The discrete wave function vector corresponding to an eigenstate is treated as the embedding vector, denoted as $\mathbf{P}^M$ where $M$ is the dimension of the embedding vector. This process can be implemented using a transformation matrix $\mathbf{G}^{N \times M}$, as follows:
\begin{equation}
    \mathbf{P}^M = \mathbf{S}^N \times \mathbf{G}^{N \times M}
\end{equation}

Each energy state $S_i$ corresponds to an energy value $e_i$, represented as $\mathbf{E}^N = \{e_1, e_2, \ldots, e_N\}$. The embedding transformation matrix $\mathbf{G}^{N \times M}$ can then be obtained through the transformation function $
TransE$. According to the Schr\"{o}dinger equation, the eigenstates $\mathbf{S}^N$ must satisfy the following three conditions:
(1) boundary condition; (2) orthogonality condition; (3) energy minimization condition.

To solve for $\mathbf{E}^N$, we define $de^N$ as a trainable weight. Then, the energy vector $\mathbf{E}^N$ can be expressed as:
\begin{equation}
    \mathbf{E}^N = e_{\min} + |de^N| \times \mathbf{I}^{N \times N}
\end{equation}
where $\mathbf{I}^{N \times N}=\begin{bmatrix}
1 & \cdots & 1 \\
\vdots & \ddots & \vdots \\
0 & \cdots & 1
\end{bmatrix}
$ is an upper triangular matrix.

The three conditions above can be formulated as follows:

\begin{enumerate}
    \item \textbf{Boundary condition:}
    
    \begin{equation}
        \sum_{i=1}^{N} |P_i^1| \to 0
    \end{equation}
    
    \item \textbf{Orthogonality condition:}
    
    \begin{equation}
        \sum_{i=2}^{N} \left|\sum_{j=1}^{M} P_i^j \times P_{i-1}^j\right| \to 0
    \end{equation}
    
    \item \textbf{Energy minimization condition:}
    
    \begin{equation}
        min\sum_{i=1}^{N} |\mathbf{de}^i|
    \end{equation}
    
\end{enumerate}

By training  $\mathbf{de}^N$ using a neural network, we can output the embedding vector $\mathbf{P}^{M}$ corresponding to the energy state $\mathbf{S}^N$, thus solving the eigenvalue problem of the Schr\"{o}dinger equation.

\subsection{\label{sec:level2b}Generator}
In the generator design of this study, multiple fully connected layers are included. These layers, where each neuron connects to all neurons in the next layer, are fundamental to neural networks. This structure enables complex nonlinear mappings between input and output, achieving high performance in solving complex problems~\cite{radhakrishnan2023mechanismfeaturelearningdeep,SCABINI2023128585,LeCun2015}. Let the input be the spatial coordinate $x \in \mathbb{R}^d$. The output of each layer is represented as a combination of a linear transformation of the input data and a nonlinear activation function. Specifically, for the output of the $l$-th layer, denoted as $h^{(l)}$, we have:
\begin{equation}
    h^{(l)}=\sigma(W^{(l)}h^{(l-1)}+b^{(l)})
\end{equation}
where $W^{(l)}$ represents the weight matrix of the layer, $b^{(l)}$ is the bias vector, and $\sigma(\cdot)$ is the activation function. $h^{(0)} = x$ denotes the input spatial coordinate.

The introduction of the activation function $\sigma(\cdot)$ enhances the nonlinear representation capability of the network. In a fully connected network, linear transformations alone are insufficient to approximate complex nonlinear mappings. By employing an activation function, the network is capable of capturing complex features more effectively. For fitting the wave functions of quantum systems, this study adopts the hyperbolic tangent (tanh) as the activation function, which is defined as:
\begin{equation}
    \sigma(x)=\tanh(x)=\frac{e^{x}-e ^{-x}}{e^{x}+e ^{-x}}
\end{equation}

The output range of the tanh function is $(-1,1)$. Its nonlinear characteristics ensure the smoothness and continuity of the wave function while providing excellent numerical stability. Compared to other activation functions, a significant advantage of the tanh function lies in its non-zero first and second derivatives, which is particularly critical when computing derivatives of the wave function.
In contrast, the commonly used Rectified Linear Unit (ReLU) activation function, while performing well in approximating linear relationships, has a significant limitation: As a piecewise linear function, ReLU has a zero second derivative across its entire domain. This makes ReLU unsuitable for tasks that require higher-order derivative computations, such as the second derivatives of wave functions in quantum mechanics. As wave functions in physical systems are typically continuous and smooth, the tanh activation function is better suited for accurately modeling the behavior of wave functions in quantum mechanics.
Additionally, the bounded output range of the tanh function ensures that variations in input amplitudes are effectively controlled, avoiding large numerical fluctuations that could negatively impact network training. This design enables the network to efficiently approximate wave functions in quantum systems, providing a solid foundation for subsequent network optimization and the incorporation of physical information.

In deep networks, increasing layer depth often leads to vanishing or exploding gradients, degrading performance. To address this issue, this study incorporates residual connections, originally proposed in ResNet by He et al.~\cite{7780459,Targ2016ResnetIR}, to stabilize training by introducing shortcut paths that ensure information flow in deep networks. Let $h^{(l)}$ denote the output of the $l$-th layer. The residual connection structure can be expressed as:
\begin{equation}
    h^{(l)}=\sigma(W^{(l)}h^{(l-1)}+b^{(l)})+h^{(l-1)}
\end{equation}
where $h^{(l-1)}$ is the output of the previous layer, directly passed to the $l$-th layer through a shortcut path and added to the nonlinearly transformed output.
This design effectively mitigates the vanishing gradient problem in deep networks. Even when the gradient information cannot propagate through the nonlinear transformation in certain layers, the shortcut path ensures the flow of information is preserved.
We incorporate multiple residual blocks after the fully connected neural network, allowing the network to maintain significant depth while stabilizing the training process and enhancing the model's representational capacity. In PINNs, where wave functions often exhibit intricate structures, residual connections enhance the model's ability to capture complex features while ensuring computational stability.

\subsection{\label{sec:level2c}Domain Decomposition}
To efficiently handle the wave functions of quantum systems, we divide the spatial domain into multiple subregions and utilize the previously described generator to each subregion. Within each subregion, the neural network generates the local wave function solution, which is then combined into a global solution using a spatial merging strategy.
Since the properties of quantum system wave functions can vary significantly across different regions--some regions exhibit sharp changes while others remain relatively smooth--this decomposition approach allows for fine-grained modeling of local complexities, avoiding redundancy in global computations~\cite{8100115,SIRIGNANO20181339,Orús2019,2005Domain}. Furthermore, domain decomposition reduces computational complexity by enabling parallel processing, thereby accelerating computation. The final merging step ensures the consistency and accuracy of the physical system, making this method highly efficient and precise for solving quantum systems.

\subsection{\textbf{Loss Function Design and Optimization Strategy}}
\subsubsection{\textbf{Loss Function Design}}
To optimize the neural network's output and accurately approximate the solution of the Schr\"{o}dinger equation, we designed a total loss function consisting of multiple components. The loss components correspond to the Schr\"{o}dinger equation residual, physical constraint, energy, and the deviation between the wave function and the embedding vector. The total loss function is defined as:
\begin{equation}
    \mathcal{L}_{\text{total}} = \lambda_s\mathcal{L}_{\text{Schr\"{o}dinger}} + \lambda_c \mathcal{L}_{\text{constraint}} +\lambda_e \mathcal{L}_{\text{energy}} +  \lambda_d \mathcal{L}_{\text{diff}}
\end{equation}
where $\lambda_s, \lambda_c, \lambda_e, \lambda_d$ are the weights representing the contributions of each loss term to the total loss. Detailed definitions and roles of these loss terms are provided below.

The residual loss of the Schr\"{o}dinger equation is the most critical component of the total loss function. It ensures the physical constraints by minimizing the deviation between the network-generated wave function $\psi(x)$ and the Schr\"{o}dinger equation. Unlike the commonly used mean squared error method, we calculate the maximum absolute value of the Schr\"{o}dinger equation residual at all sampled points $x_i$. This approach prevents excessively large residuals at any sampled point. The specific form of this loss term is:

\begin{equation}
    \mathcal{L}_{\text{Schr\"{o}dinger}} = \max_{i=1, \dots, M} \left| -\frac{\hbar^2}{2m} f_{PPi} + V_i\psi_i-E\psi_i \right|
\end{equation}
where the optimization goal is to control the maximum error, rather than simply reducing the average error. This method ensures that the network's output satisfies the physical equation uniformly across all sampled points, avoiding the common issue in MSE-based approaches where large errors at a few points can dominate and distort the optimization process.
In training, the Schr\"{o}dinger equation residual loss is assigned a weight of  $\lambda_{s} = 1$, giving it the highest importance in the total loss function. This is because the physical constraints imposed by the Schr\"{o}dinger equation are crucial for the correctness of the wave function.

The physical constraint loss ensures the validity of the embedding vectors produced by the embedding layer, encompassing both boundary conditions and orthogonality constraints. According to the energy transformation function $TransE$, any energy value can theoretically generate an embedding vector in the absence of boundary conditions and orthogonality constraints. However, to ensure the correctness of the solution, both of these conditions are essential. Therefore, the physical constraint loss is designed to enforce these requirements, and is defined as:
\begin{equation}
   \mathcal{L}_{constraint} = \sum_{i=1}^{N} \left|P_i^1\right| + \sum_{i=2}^{N} \left| \sum_{j=1}^{M} P_i^j \times P_{i-1}^j \right| \to 0
\end{equation}
Where, $\mathcal{L}_{constraint}$ represents the boundary and orthogonality terms, with the first term ensuring the boundary condition is satisfied and the second term ensuring orthogonality between wave functions. This loss is assigned a weight of $\lambda_c = 1$, as it should theoretically converge to $0$. Additionally, physical constraint are the fundamental requirement in physical systems and cannot be neglected.

The energy loss guides the optimization of energy values. The Schr\"{o}dinger equation is a multi-solution problem, and energy minimization guides the network to satisfy these conditions with the smallest possible energy. The energy loss is defined as:

\begin{equation}
    L_{energy} = \sum_{i=1}^{N} \left|2 \cdot \frac{e^{e_i}}{e^{e_i} + e^{-e_i}}\right|
\end{equation}

Theoretically, energy values can take any real number, and the principle of energy minimization is a relative concept. To constrain the energy loss within a bounded range, we apply the above transformation function. The weight of the energy loss term is set to $\lambda_{e} = 0.001$.

The $\mathcal{L}_{\text{diff}}$ is designed to assist in the generation of wave functions. The wave functions generated by the network belong to result-based data and are directly optimized to satisfy the Schr\"{o}dinger equation, making the training process highly challenging. To address this, the network employs an embedding layer to generate process-based data to aid in optimizing the wave function. This is achieved by ensuring a certain level of consistency between the wave function and the embedding vector. The loss is defined as:

\begin{equation}
    \mathcal{L}_{\text{diff}} =\frac{1}{M} \sum_{i=1}^M \left(e^{\alpha\left|\psi_i-P_i\right|}\right)^2
\end{equation}
where $\alpha$ represents the degree of consistency between the wave function and the embedding vector, typically set to $\alpha=4$. Although $\mathcal{L}_\text{diff}$ is important, it primarily serves as an auxiliary term. Its weight is generally set to $\lambda_d=0.01$, ensuring that the network receives sufficient guidance during the initial training phase without interfering with other physical constraints.

\subsubsection{\textbf{Optimization Strategy}}
To efficiently solve the Schr\"{o}dinger equation with neural networks while ensuring stable training, we adopted an optimization strategy based on the Adam algorithm, enhanced with additional techniques to improve performance.

We started with the Adam optimizer~\cite{Kingma2014AdamAM}, initialized with a learning rate of 0.001. The Adam optimizer is well suited for complex nonlinear optimization tasks, as it handles sparse gradients and dynamically adjusts learning rates, offering both stability and reliable convergence.
To further refine the training process, we utilized the ReduceLROnPlateau technique, which gradually reduces the learning rate by a factor of $0.9$ when the training loss shows no significant decrease. This strategy enables the model to maintain steady convergence, particularly during the later stages of training~\cite{7926641}.

We also employed an early stopping mechanism~\cite{Prechelt2012} to prevent overfitting and save computational resources. Training stops automatically if no substantial improvement in loss is observed over 1000 epochs, ensuring efficient use of time.

Finally, we implemented model checkpointing~\cite{Heaton2018} to preserve the best-performing model during training. This system saves the model weights whenever the training loss reaches a new minimum. If the loss fluctuates in later epochs, checkpointing ensures that we can revert to the optimal state.

\section{\label{sec:level1c}Results}
\subsection{\textbf{Infinite Potential Well}}
The one-dimensional infinite potential well is a classical quantum mechanics problem with well-defined analytical solutions for wave functions and energy eigenvalues. The potential function  $V(x)$ is infinite outside the specified interval and zero within it, given by:
\begin{equation}\label{eq:well}
    V(x) =
\begin{cases} 
0, & x_{min} \leq x \leq x_{max}, \\
\infty, & \text{otherwise}.
\end{cases}
\end{equation}

To simplify calculations and formula representation, this study adopts natural units: $\hbar = 1,\; m = 1$. The domain is set to $[-4,4]$, with the analytical solutions for wave functions and energy eigenvalues given by:
\begin{equation}
    \psi_n(x) = \frac{1}{2} \sin\left(\frac{(n+1) \pi}{8} (x + 4)\right), \quad n = 0, 1, 2, \dots,
\end{equation}
\begin{equation}
    E_n = \frac{(n+1)^2 \pi^2}{128}, \quad n = 0, 1, 2, \dots.
\end{equation}

\begin{figure*}
    \centering
    \includegraphics[width=1.0\linewidth]{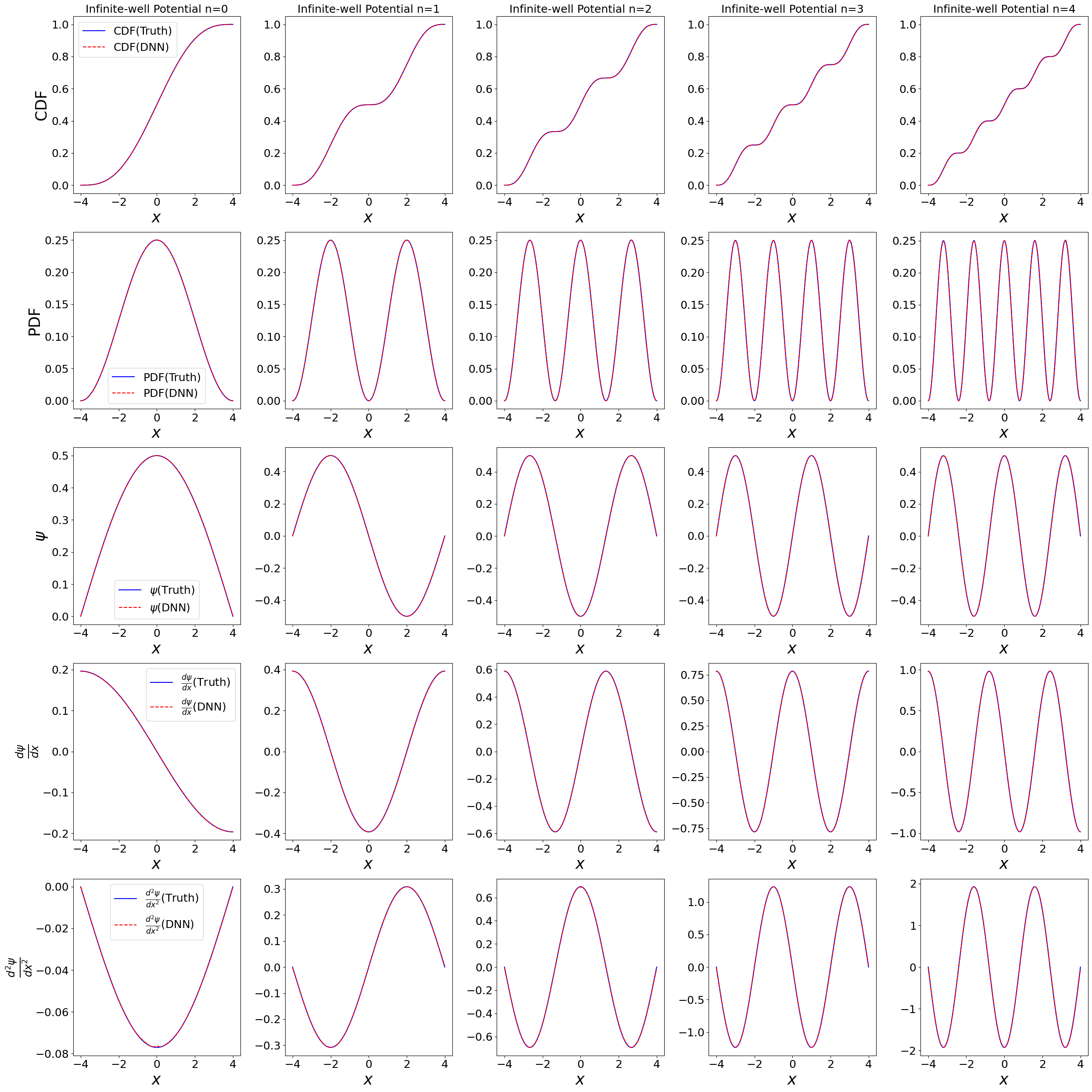}
    \caption{Results of the neural network for the infinite potential well, Eq.~\ref{eq:well}. Each column represents a quantum state $(n=0$ to $n=4)$, with subplots comparing theoretical (solid blue curve) and neural network results (dashed red curve) for the CDF, PDF, wavefunction, and its derivatives, demonstrating high accuracy across all states.}
    \label{fig:Infinite-well-2}
\end{figure*}

FIG.~\ref{fig:Infinite-well-2} shows the results of the neural network proposed for the one-dimensional infinite potential well. Each column corresponds to a different eigenstate, with quantum numbers $n=0, 1, 2, 3, 4$. Each column includes five subplots comparing the following:
\begin{enumerate}
    \item the cumulative distribution function (CDF) of the wave function,
    \item the probability density function (PDF) of the wave function,
    \item the wave function,
    \item the first derivative of the wave function, 
    \item the second derivative of the wave function.
\end{enumerate}
In the figure, blue solid lines represent analytical solutions, while the red dashed lines show the neural network results.

\begin{figure}
    \centering
    \includegraphics[width=0.8\linewidth]{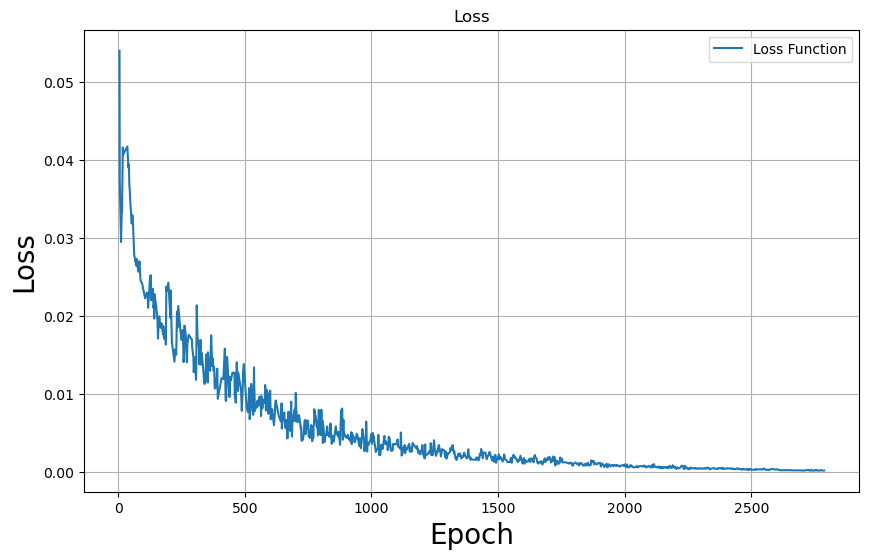}
    \caption{The figure demonstrates the rapid convergence of the loss function during the training process for the ground state of the infinite potential well. The loss function significantly decreases within the first 500 epochs, indicating efficient optimization. The steady decline and stabilization in subsequent epochs highlight the model's ability to achieve high accuracy with minimal training iterations.}
    \label{fig:loss-}
\end{figure}

To thoroughly evaluate the accuracy of the wave functions generated by the neural network, we performed a comparative analysis from multiple perspectives. Specifically, the analysis focused on five aspects: the cumulative distribution function (CDF), the probability density function (PDF), the wave function, and its first and second derivatives. The first and second derivatives highlight the continuity and smoothness of the wave function. The CDF illustrates its distribution in the normalized space, indirectly capturing cumulative errors, while the PDF reveals its spatial probability distribution.  The results show that all neural network-generated curves align closely with the analytical solutions.

To further quantify the wave function error, we introduced the normalized wavefunction fidelity $(F)$ to measure the similarity between two solutions. The fidelity is defined as:
\begin{equation}
    F(\psi_{\text{1}}, \psi_{\text{2}}) = \frac{\left| \int_{x_{\text{min}}}^{x_{\text{max}}} \psi_{\text{1}}^*(x) \psi_{\text{2}}(x) dx \right|^2}{\left( \int_{x_{\text{min}}}^{x_{\text{max}}} |\psi_{\text{1}}(x)|^2 dx \right) \left( \int_{x_{\text{min}}}^{x_{\text{max}}} |\psi_{\text{2}}(x)|^2 dx \right)}.
\end{equation}
where $\psi_{\text{1}}(x)$ is the wave function generated by the neural network, while $\psi_{\text{2}}(x)$ denotes the analytical solution. A fidelity $F$ close to 1 indicates strong agreement between the neural network solution and the analytical solution. The relevant data is shown in Table \ref{tab:InfiniteWellData}.

\begin{table}[h!]
\centering
\begin{ruledtabular}
\begin{tabular}{cccccc}
\textbf{State} & $F$ & $E_\text{error}~(\%)$ & $E_\text{NN}$ & $E_\text{Analytical}$ \\
\hline
$n=0$ & 0.99999994 & 0.2239 & 0.07693 & 0.07711 \\
$n=1$ & 0.99999982 & 0.1883 & 0.30784 & 0.30843 \\
$n=2$ & 0.99999992 & 0.1943 & 0.69261 & 0.69396 \\
$n=3$ & 0.99999989 & 0.1955 & 1.23129 & 1.23370 \\
$n=4$ & 0.99999980 & 0.1924 & 1.92395 & 1.92766 \\
\end{tabular}
\end{ruledtabular}
\caption{Comparison of wavefunction fidelity (\(F\)), energy error (\(E_\text{error}\)), neural network energy (\(E_\text{NN}\)), and analytical energy (\(E_\text{Analytical}\)) for different quantum states in the infinite potential well.}
\label{tab:InfiniteWellData}
\end{table}

We compare our proposed method with the PINN-based approach for solving the ground state of one-dimensional quantum systems to that described in Ref.~\cite{CPC:10.1088/1674-1137/acc518}, which also employs PINNs and evaluates the ground state energy error and wavefunction fidelity. The comparison of energy error and fidelity between the proposed method and the reference method are summarized in Table\ref{tab:Comparison1}.

\begin{table}[h!]
\centering
\begin{ruledtabular}
\begin{tabular}{ccccc}
\textbf{State} & $E_\text{error}$ \textbf{(Ours) (\%)} & $E_\text{error}$ \textbf{(Ref) (\%)} & $F$ \textbf{(Ours)} & $F$ \textbf{(Ref)} \\
\hline
Ground State   & 0.2239                   & 0.9987                    & 0.99999994        & 0.9977475            \\
\end{tabular}
\end{ruledtabular}
\caption{Comparison of energy error (\(E_\text{error}\)) and wavefunction fidelity (\(F\)) between our proposed method and the reference method~\cite{CPC:10.1088/1674-1137/acc518} for the ground state.}
\label{tab:Comparison1}
\end{table}

As shown in Table \ref{tab:Comparison1}, our proposed method achieves a significantly lower energy error of $0.2239\%$ for the ground state, compared to $0.9987\%$ for the reference method. Moreover, the wavefunction fidelity of the proposed method, at $0.99999994$, is closer to $1$ compared to the fidelity of $0.9977475$ for the reference method. This accuracy improvement is attributed to the architectural design of the proposed method. The introduction of the embedding layer and specific spatial processing modules equips the network to better handle boundary conditions, enabling faster convergence to the physical solution during training and enhancing both precision and stability as shown in Fig.\ref{fig:loss-}.

\subsection{\textbf{Other One-Dimensional Potentials}}
To evaluate the generalization capability of the proposed method, we applied the same network parameters as those for the infinite potential well to solve three additional potentials: the harmonic oscillator, Woods-Saxon, and double-well potentials.
\begin{itemize}
\item[(a)] 
The harmonic oscillator potential is given by:
\begin{equation}
    V(x)=\frac{1}{2}m\omega^2x^2
\end{equation}
where $\omega$ represents the angular frequency, and $m$ denotes the mass, both of which are simplified to 1.
\item[(b)]
The Woods-Saxon potential is expressed as:
\begin{equation}
    V(x) = \frac{V_0}{1 + \exp\left(\frac{x - R_0}{a_0}\right)}
\end{equation}
where $V_0=-1$, $R_0=6.2$, and $a_0=0.1$ are parameters representing the potential depth, potential radius, and surface thickness, respectively.
\item[(c)]
The double-well potential takes the form:
\begin{equation}
    V = -e^{-h(x + l - a)(x + l + a)} - e^{-h(x - l + a)(x - l - a)}
\end{equation}
where $h=4,a=0.5,l=1.5$ control the depth, position, and width of the double-well.
\end{itemize}
The analytical solutions for the one-dimensional harmonic oscillator are:
\begin{equation}
    \psi_n(x) = (-1)^n \cdot \pi^{-1/4} \cdot \sqrt{\frac{n!}{2^n}} \cdot H_n(x) \cdot e^{-x^2 / 2}, \quad n = 0, 1, 2, \dots,
\end{equation}
\begin{equation}
    E_n = 0.5 + n, \quad n = 0, 1, 2, \dots,
\end{equation}
where $H_n(x)$ is the Hermite polynomial and $n$ is the quantum number.
 As the Woods-Saxon and double-well potentials lack analytical solutions, we use high-precision matrix eigenvalue methods as reference solutions to evaluate the neural network's performance.

\begin{figure*}[t]
    \centering
    \begin{subfigure}{\textwidth}
        \centering
        \includegraphics[width=\textwidth]{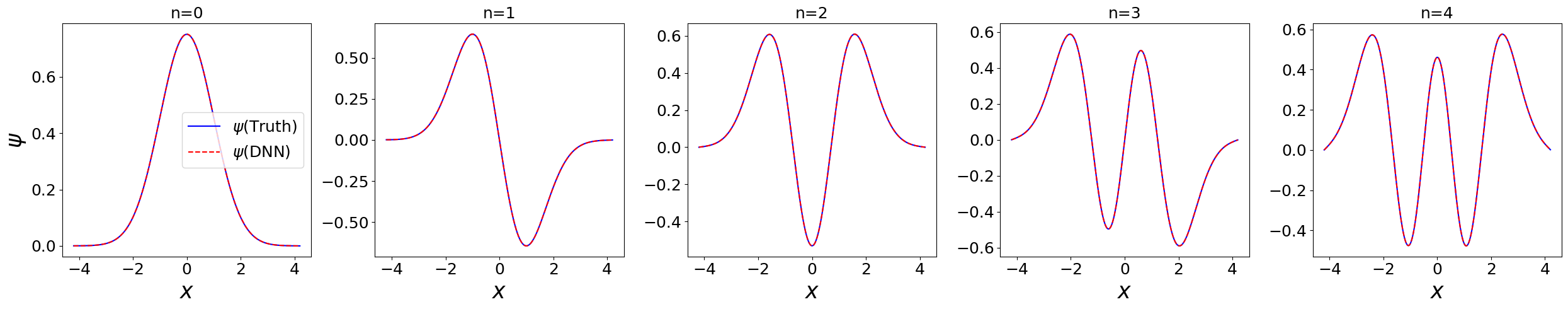}
        \caption{Harmonic Oscillator}
        \label{fig:subfig1-2}
    \end{subfigure}
    \vfill
    \begin{subfigure}{\textwidth}
        \centering
        \includegraphics[width=\textwidth]{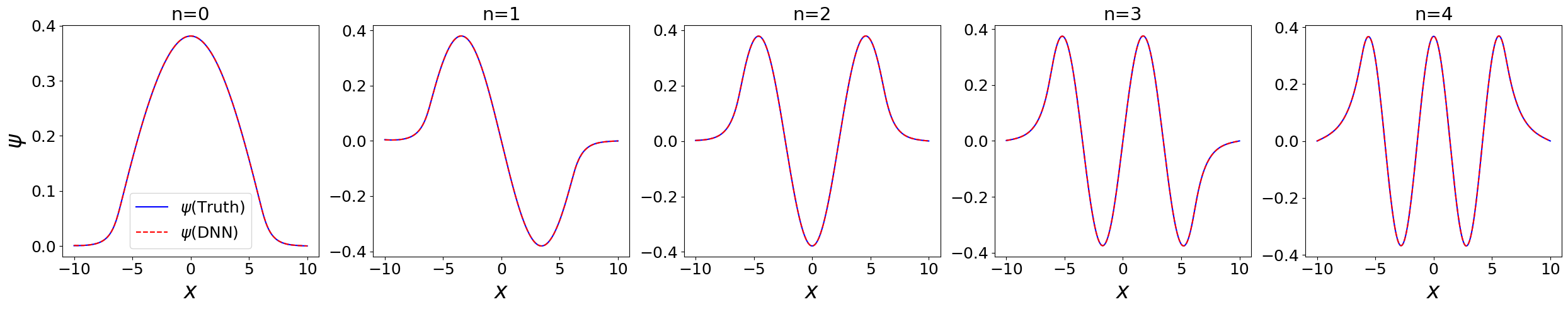}
        \caption{Woods-Saxon Potential}
        \label{fig:subfig2-2}
    \end{subfigure}
    \vfill
    \begin{subfigure}{\textwidth}
        \centering
        \includegraphics[width=\textwidth]{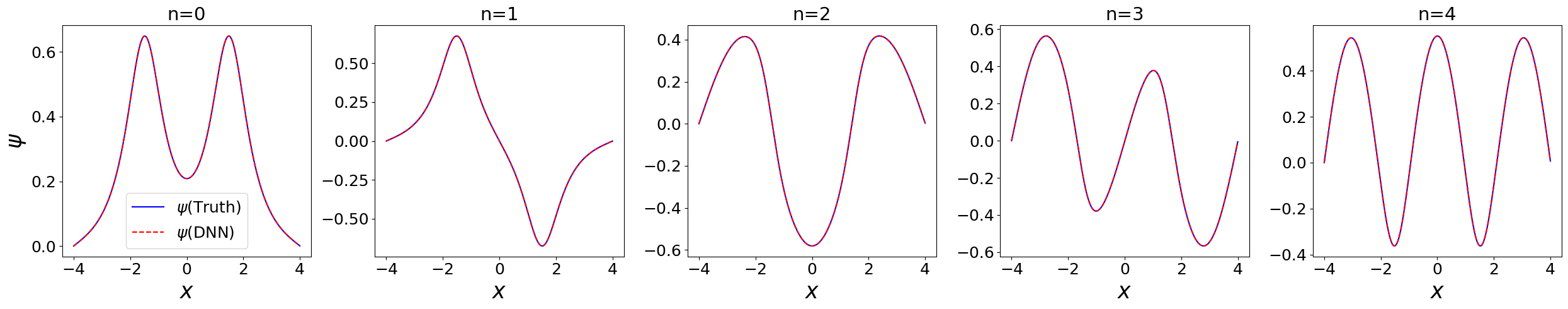}
        \caption{Double-Well Potential}
        \label{fig:subfig3-2}
    \end{subfigure}
    \caption{Comparison of our results (dashed red curve) to theoretical solutions (solid blue curve) for different potentials: (a) Harmonic Oscillator, (b) Woods-Saxon Potential, (c) Double-Well Potential.}
    \label{fig:three_potentials}
\end{figure*}

FIG.~\ref{fig:three_potentials} shows the results obtained by the proposed neural network  for the harmonic oscillator, Woods-Saxon, and double-well potentials. Each column corresponds to a quantum state with quantum numbers $n = 0, 1, 2, 3, 4$.
The wave functions in FIG.~\ref{fig:three_potentials} are generated using the same network architecture and hyperparameters and closely align with the theoretical solutions for all three potential systems. This demonstrates the proposed network's ability to capture the fundamental physical properties of quantum state wave functions. It validates the method's effectiveness and accuracy across different potential systems, highlighting its strong generalization capability.

Table~\ref{tab:potentials_comparison} summarizes the fidelity and energy error for the harmonic oscillator, Woods-Saxon, and double-well potentials. Table~\ref{tab:energy_error_comparison} compares the results for the one-dimensional harmonic oscillator with those from a neural network solver based on the variational method in Ref.~\cite{PhysRevA.103.032405}.
As shown in Table \ref{tab:energy_error_comparison}, the proposed method achieves significantly smaller errors for the first five eigenenergies compared to Ref.~\cite{PhysRevA.103.032405}.

\begin{table*}[h!]
    \centering
    \begin{ruledtabular}
    \begin{subtable}[t]{0.3\textwidth}
        \centering
        \caption{Harmonic Oscillator}
        \begin{tabular}{ccc}
            \textbf{State} & \textbf{$E_\text{error}~(\%)$} & \textbf{F} \\
            \hline
            $n=0$ & 0.0182 & 0.999999970 \\
            $n=1$ & 0.0015 & 0.999999934 \\
            $n=2$ & 0.0011 & 0.999999868 \\
            $n=3$ & 0.0104 & 0.999999862 \\
            $n=4$ & 0.0626 & 0.999999659 \\
        \end{tabular}
    \end{subtable}
    \hfill
    \begin{subtable}[t]{0.3\textwidth}
        \centering
        \caption{Woods-Saxon Potential}
        \begin{tabular}{ccc}
            \textbf{State} & \textbf{$E_\text{error}~(\%)$} & \textbf{F} \\
            \hline
            $n=0$ & 0.00032 & 0.99999982 \\
            $n=1$ & 0.00031 & 0.99999987 \\
            $n=2$ & 0.00027 & 0.99999992 \\
            $n=3$ & 0.00344 & 0.99999974 \\
            $n=4$ & 0.01370 & 0.99999983 \\
        \end{tabular}
    \end{subtable}
    \hfill
    \begin{subtable}[t]{0.3\textwidth}
        \centering
        \caption{Double-Well Potential}
        \begin{tabular}{ccc}
            \textbf{State} & \textbf{$E_\text{error}~(\%)$} & \textbf{F} \\
            \hline
            $n=0$ & 0.0046 & 0.99999990 \\
            $n=1$ & 0.0216 & 0.99999990 \\
            $n=2$ & 0.1477 & 0.99999995 \\
            $n=3$ & 0.2724 & 0.99999989 \\
            $n=4$ & 0.2471 & 0.99999978 \\
        \end{tabular}
    \end{subtable}
    \end{ruledtabular}
    \caption{Fidelity ($F$) and energy error ($E_\text{error}$) for different potentials. Each subtable corresponds to a specific potential system.}
    \label{tab:potentials_comparison}
\end{table*}

\begin{table}[h!]
\centering
\begin{ruledtabular}
\begin{tabular}{cccccc}
\textbf{State} & $n=0$ & $n=1$ & $n=2$ & $n=3$ & $n=4$ \\
\hline
Ours (\%)  & 0.0182 & 0.0015 & 0.0011 & 0.0104 & 0.0626 \\
Ref (\%) & 0.21   & 0.30   & 0.52   & 0.63   & 0.20   \\
\end{tabular}
\end{ruledtabular}
\caption{Energy error comparison between our proposed method and the reference method~\cite{PhysRevA.103.032405}.}
\label{tab:energy_error_comparison}
\end{table}

\subsection{\textbf{The Radial Coulomb Potential}}

To assess the extensibility of the proposed method, we applied it to the radial Coulomb potential equation. The radial Coulomb potential equation differs from the one-dimensional Schr\"{o}dinger equation in form and normalization, making it an ideal test case for validating the method's adaptability. For simplicity in calculations and expressions, we adopted  \( V(r) = -\frac{1}{r} \) and Coulomb potential in a simplified unit system. 

 The time-independent Schr\"{o}dinger equation in spherical coordinates $(r, \theta, \phi)$ takes the form:

\begin{equation}
    -\frac{1}{2} \nabla^2 \Psi(r, \theta, \phi) + V(r) \Psi(r, \theta, \phi) = E \Psi(r, \theta, \phi)
\end{equation}
where $ \nabla^2 $ is the Laplacian operator, $ V(r) $ is the potential energy, $\Psi(r, \theta, \phi)$ is the wave function, and $ E $ is the energy.
Exploiting the spherical symmetry of the system, the wave function is separated into radial and angular parts as:
\begin{equation}
\Psi(r, \theta, \phi) = \psi_{n,l}(r) Y_{l,m}(\theta, \phi)
\end{equation}
where $\psi_{n,l}(r)$ is the radial part of the wave function, $Y_{l,m}(\theta, \phi)$ are the spherical harmonics corresponding to the angular momentum quantum numbers $l$ and $m$. Focusing on the radial part $\psi_{n,l}(r)$, we consider the $l = 0$ case and replace
$r$ with $x$ for notational simplicity.

The corresponding radial Schr\"{o}dinger equation is:
\begin{equation}
    \frac{1}{2}(\frac{d^2\psi(x)}{dx^2} + \frac{2d\psi(x)}{x dx}) - \frac{1}{x}\psi(x) = E\psi(x)
\end{equation}

The analytical solution can be expressed as:
\begin{equation}
    \psi_n(x) = \sqrt{\left(\frac{2}{n+1}\right)^3(n!)}\frac{1}{\sqrt{8\pi(n+1)((n+1)!)}}e^{-\frac{x}{n+1}}L_n^1(\frac{2x}{n+1}), \quad n=0,1,2,\dots
\end{equation}

\begin{equation}
    E_n = -\frac{1}{2(n+1)^2}, \quad n=0,1,2,\dots
\end{equation}
where $L_n^1(x)$ is the generalized Laguerre polynomial, and $n$ is the quantum number. Certain adjustments were made to the network to handle the radial Coulomb potential equation. The key modifications include:
\begin{enumerate}
    \item \textbf{Schr\"{o}dinger Equation Loss Term}:
    \begin{equation}
        \mathcal{L}_{\text{Schr\"{o}dinger}} = \max_{i=1, \dots, M} \left| -\frac{1}{2} \left(f_{PPi} + \frac{1}{x_i}f_{Pi}\right) + \frac{1}{x_i}\psi_i - E\psi_i \right|
    \end{equation}

    \item \textbf{In the Embedding Layer}:
    \begin{equation}
        P_{i+2} = 2\left(1 - \frac{\Delta x}{x_i}\right)P_{i+1} + \left(2(\Delta x)^2(V_i - e) + \frac{2\Delta x}{x_i} - 1\right)P_i
    \end{equation}

    \item \textbf{Normalization Formula}:
    \begin{equation}
        \sum_{i=1}^{M} 4\pi x_i^2(P_i)^2 \Delta x \approx 1
    \end{equation}

    \item \textbf{Boundary Condition Loss Term}:
    \begin{equation}
        L_{\text{constraint}} = \sum_{i=1}^{N} \left|P_i^1 - P_i^2\right| + \sum_{i=2}^{N} \left|\sum_{j=1}^{M} P_i^j \times P_{i-1}^j\right|
    \end{equation}
\end{enumerate}

\begin{figure}[H]
    \centering
    \includegraphics[width=1.0\linewidth]{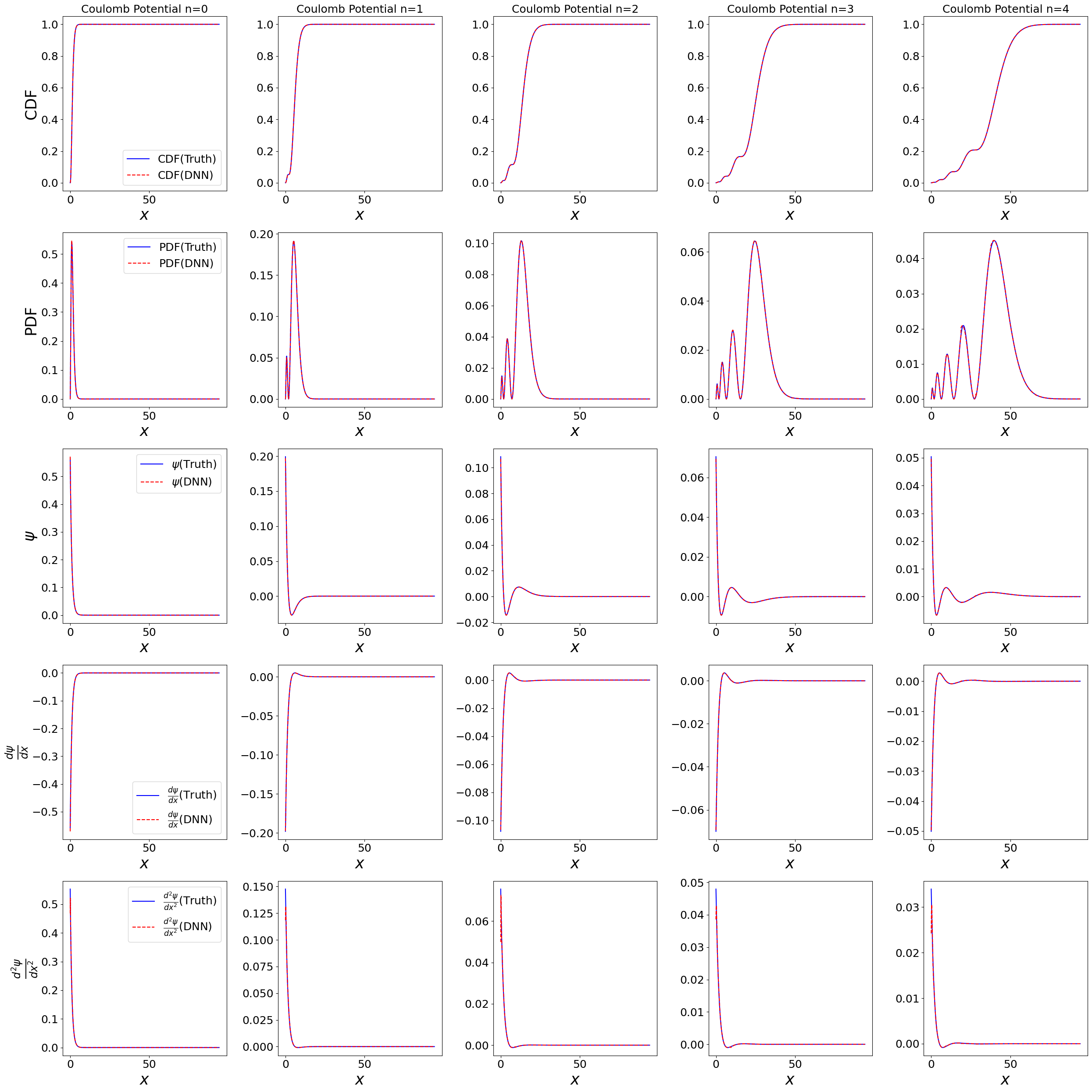}
    \caption{Results of the neural network for the radial Coulomb potential. Subplots compare the cumulative distribution function (CDF), probability density function (PDF), wavefunction, first derivative, and second derivative between theoretical solutions (solid blue curve) and neural network predictions (dashed red curve) for quantum states $n=0,1,2,3,4$.}
    \label{fig:Coulomb-2}
\end{figure}

FIG.~\ref{fig:Coulomb-2} shows the results obtained by the proposed neural network for the radial Coulomb potential. Each column represents a different quantum state, with quantum numbers $n=0,1,2,3,4$. Each subplot compares the cumulative distribution function (CDF), probability density function (PDF), wave function, first derivative, and second derivative.

\begin{table}[h!]
\centering
\begin{ruledtabular}
\begin{tabular}{ccccc}
\textbf{States} & \textbf{F} & $E_\text{error}~(\%)$ & $E_\text{NN}$ & $E_\text{Analytical}$ \\
\hline
$n=0$ & 0.99999971 & 1.7361 & -0.49132 & -0.50000 \\
$n=1$ & 0.99999913 & 0.0660 & -0.12508 & -0.12500 \\
$n=2$ & 0.99999866 & 1.3054 & -0.05483 & -0.05556 \\
$n=3$ & 0.99999745 & 1.5335 & -0.03077 & -0.03125 \\
$n=4$ & 0.99998201 & 0.9221 & -0.02018 & -0.02000 \\
\end{tabular}
\end{ruledtabular}
\caption{Comparison of wavefunction fidelity (\(F\)), energy error (\(E_\text{error}\)), neural network energy (\(E_\text{NN}\)), and analytical energy (\(E_\text{Analytical}\)) for different quantum states in the radial Coulomb potential.}
\label{tab:Coulomb data}
\end{table}

Table \ref{tab:Coulomb data} summarizes the results obtained by the neural network for different quantum states in the radial Coulomb potential. The data includes the wavefunction fidelity ($F$), energy error ($E_\text{error}$), neural network-predicted energy ($E_\text{NN}$), and analytical energy ($E_\text{Analytical}$). The results show that for quantum numbers $n = 0, 1, 2, 3, 4$, the neural network-generated wavefunctions exhibit high fidelity (close to $1$), with relatively low energy errors. This indicates that the network effectively captures the quantum state properties under the radial Coulomb potential.

The results show that our network achieves high accuracy in solving the radial Coulomb potential. This further confirms that the proposed method extends beyond one-dimensional time-independent Schr\"{o}dinger equations, demonstrating adaptability to other equation types. This generality arises from the core design of the network, particularly the introduction of the embedding technique: by generating essential intermediate (process-driven) data, the embedding technique effectively alleviates the challenges of directly solving constrained optimization problems, enabling the network to approximate the target solution more rapidly and stably. In addition, the generator, composed of fully connected layers and residual layers, produces flexible functional curves that are optimized to satisfy the constraints of the target equation. According to our representation principle, this network architecture functions effectively when the discretized target equation can be represented by a small initial set that describes the entire result set. This highlights our core idea: combining the intrinsic properties of equations with the embedding technique to generate process-driven data simplifies the training process and enhances the network's generalization capability.

\section{\label{sec:level1d}Summary}
This study proposed a novel physics-informed neural network (PINN) for solving ground- and excited-state wavefunctions in one-dimensional quantum systems. 
The generator network comprises fully connected and residual layers. The residual network effectively approximates the true wavefunctions, enhancing their accuracy significantly. To overcome the limitations of traditional PINN methods in handling process-driven data, we introduced an embedding layer to capture the relationship between energy and potential systems. This design simplifies training and improves the network's generalization capability. Experimental results show that the proposed method achieves high accuracy across various potential systems, including the infinite potential well, harmonic oscillator, Woods-Saxon, and double-well potentials.

By leveraging the intrinsic properties of the Schr\"{o}dinger equation, the proposed method demonstrates strong extensibility. Test results indicate that, with appropriate adjustments, the method achieves high accuracy in solving the radial Coulomb potential equation, which differs in form from the one-dimensional Schr\"{o}dinger equation. This suggests that the proposed method extends beyond the one-dimensional Schr\"{o}dinger equation, offering a novel framework for addressing PDEs. Under certain conditions, this method can be extended to other forms of PDEs.

However, the complex coupling interactions in high-dimensional quantum systems prevent the method from being directly applied to higher-dimensional cases. Future work will focus on exploring wavefunction representations in high-dimensional spaces and incorporating additional physical constraints. This will enable the method to extend to higher dimensions and more complex quantum many-body problems, broadening its applications in quantum mechanics and beyond.
\section{Acknowledgments}
This work is supported by the National Natural Science Foundation of China (Grants No. 12035006). Computations are performed at the High-Performance Computing Center of Wuhan University of Science and Technology. Hanlin Li and Yi-Qiang Wu thank Prof. Long-Gang Pang for helpful discussions for this work.

\bibliography{apssamp}

\end{document}